# DYNAMICAL IMPLICATIONS OF ADJUSTMENTS TO PROPER TIME CAUSED BY HIGHER DIMENSIONS: A NOTE

Paul S. Wesson

Dept. Physics and Astronomy, University of Waterloo, Waterloo, Ontario N2L 3G1, Canada.

Abstract: When the proper time of general relativity is adjusted to reflect the possible existence of higher dimensions, small dynamical effects appear in spacetime of the sort usually associated with the cosmological constant, Hubble's Law and Heisenberg's relation.

Key Words: Higher dimensions, Cosmological constant, Hubble's Law, Heisenberg's relation

Correspondence: Mail to Waterloo (above); email = psw.papers@yahoo.ca

1. <u>Introduction</u>

Proper time is the prime parameter used to describe relativitstic physics. For example, causality in spacetime is commonly defined in terms of the element of proper time by $ds^2 \geq 0$. However, in unified theories where there are more than 4 dimensions, the corresponding quantity is given by the ND interval, and in general is $dS \neq ds$. That is, there is a mismatch between the natural elements of proper time in 4D and ND. This is usually taken to be small, because the dynamical effects of the extra dimensions are apparently small. This situation is compounded by a related and more significant issue: in order to make contact with extant knowledge, dynamical consequences of ND theories are frequently presented in terms of the 4D proper time $s$ rather than the higher-dimensional $S$. Practically, this is understandable because $S$ cannot be evaluated directly. Theoretically, however, it is quite possible that $dS/ds \neq 1$, and that physical processes proceed at different rates in ND and 4D as measured by their intrinsic proper times. An illustration is provided by the motion of a test particle in 5D Minkowski space [1, 2]: the velocity is constant as measured in $S$ but accelerated when measured in $s$. The difference is due to the mismatch between the 5D and 4D 'proper' times. In this note, the subject will be given a brief examination. The results are interesting: the algebraic mismatch between 5D and 4D dynamics tends to mimic physical effects which are already familiar from cosmology and quantum mechanics. The main result concerns the cosmological 'constant' $\Lambda$ in Einstein's theory of general relativity, which can be understood as the consequence of an acceleration between the 4D and 5D reference frames.

A 5D model is used, as the simplest extension of 4D general relativity. The line element is $dS^2 = g_{AB}dx^A dx^B (A, B = 0,123,4)$ and smoothly embeds the spacetime line element $ds^2 = g_{\alpha\beta}dx^\alpha dx^\beta (\alpha, \beta = 0,123)$. The 5D manifold is noncompactified, as in membrane theory or space-time-matter theory [3, 4]. Various technical results from the literature will be used below, which ultimately rest on Campbell's embedding theorem [5, 6]. Particular use will be made of the classical account by Rindler [7] on the embedding of the local deSitter solution of 4D general relativity in 5D Minkowski space, which showed that deS$_4$ may be viewed as a pseudosphere in $M_5$. A related account is the older one of Dirac [8], which was applied to the properties of particles. However, earlier workers were apparently unaware that a better way of embedding solutions of Einstein's 4D equations in 5D is offered by canonical space [9, 10]. This has a quadratic factor in the extra coordinate multiplied onto 4D spacetime, plus an extra flat dimension. The general case ($C_5$) can embed *any* 4D solution; while the 'pure' case where there is no dependence of spacetime on the extra coordinate ($C_5^*$) provides a convenient way to embed all 4D *vacuum* solutions, including Schwarzschild-deSitter. Since 5D canonical space is algebraically broad and can accommodate all observational data [2, 10, 11] it will be used below.

2. <u>5D Embedding and 4D Physics</u>

Below, the fundamental constants will be absorbed by a choice of units, unless they are needed to make a physical point. The coordinates are chosen to be $x^0 = t$, $x^{123} = xyz$ or $r\theta\phi$ and $x^4 = \ell$. (This last should not necessarily be identified with the distance from the singular hypersurface in membrane theory, or the geometrical measure of particle mass in space-time-matter theory.) The extra dimension can be space-like or time-like, and there is no problem with closed time-like paths because the extra coordinate does not have the physical meaning of a time, being rather connected to the properties of matter.

When spacetime is extended by an extra dimension, in general the 4D metric tensor can depend on the extra coordinate, via $g_{\alpha\beta} = g_{\alpha\beta}(x^\gamma, \ell)$. The 4-velocities may be normalized as usual, via $g_{\alpha\beta}(x^\gamma, \ell) u^\alpha u^\beta = 1$ or 0 for massive and massless test particles. Here it is conventional to define $u^\alpha \equiv dx^\alpha/ds$, using the 4D proper time. However this procedure is in general *not* equivalent to using the full 5D metric, with $g_{AB} = g_{AB}(x^\gamma, \ell)$, $g_{AB} U^A U^B = 1$ or 0 and $U^A \equiv dx^A/dS$. Notably, in general $dS/ds \neq 1$. This mismatch is connected to the appearance in 4D of an extra force per unit mass or acceleration [1, 2, 10]:

$$f^\mu = \left( -\frac{1}{2} \frac{\partial g_{\alpha\beta}}{\partial \ell} u^\alpha u^\beta \right) \frac{d\ell}{ds} u^\mu \quad . \tag{1}$$

This force has been noted in the literature in various contexts, and has been interpreted in different ways. For example, the fact that (1) acts *parallel* to the 4-velocity implies that it could be indicative of a slow change in the rest mass of a test particle *m* moving with 3D velocity *v*, because if momentum is conserved with $d(mv)/ds = 0$ then $dv/ds = -(v/m)(dm/ds)$. However, no mass parameter appears in (1); and in any case the force is clearly due to the relative motion of the 4D and 5D frames as measured by $d\ell/ds$, and is coupled to the 4D frame via the scalar term $(\partial g_{\alpha\beta}/\partial\ell)u^\alpha u^\beta$. That is, the force is inertial in the Einstein sense. Nevertheless, it is in principle measurable, though its effects in practice are expected to be small.

Minkowski space in 5D shows the acceleration (1) if motion is measured in terms of $s$ rather than $S$ [1]. Investigation shows that the motion is either steady or accelerated, depending on whether it is measured in terms of $S$ or $s$. With appropriately-defined constants, the motions in the two frames are given by

$$dx^A/dS = \text{constants} \tag{2.1}$$

$$x^A = x_0^A e^{\pm s/L} \tag{2.2}$$

In the latter relation, $L \equiv (1/\ell)(d\ell/ds)$ is a constant involving the extra coordinate, and the sign choice reflects the reversibility of the motion in the extra dimension.

The cosmological constant $\Lambda$, or at any rate a parameter like it, enters the picture if (2.2) is applied to the radial direction $x^A = x^2 = r$. The equation of motion can be written

$$\frac{d^2 r}{ds^2} = \frac{r}{L^2} \equiv \frac{\Lambda r}{3}, \qquad \Lambda \equiv \frac{3}{L^2}. \tag{3}$$

This means that for a collection of test particles which exists in 5D but is interpreted in terms of 4D proper time, there is a force on any one of them which has the same form as that usually attributed to the cosmological constant in Einstein's theory.

This might, of course, be regarded as merely an algebraic accident. However, it should be noted that the first derivative of (2.2) gives the 3D radial velocity as $v = Hr$ where $H \equiv 1/L$, which is recognized as Hubble's law. These dynamical coincidences suggest a more accurate investigation, using not $M_5$ (in which parameters like $\Lambda$ and $H$ are really foreign) but $C_5$, the canonical space of 5D general relativity.

Canonical coordinates in 5D relativity lead to certain algebraic simplifications, analogous to how in ordinary 3D space some problems are better couched in spherical polar coordinates than Cartesian coordinates [9]. The general form is

$$dS^2 = (\ell/L)^2 g_{\alpha\beta}(x^\gamma, \ell) \pm d\ell^2 \quad , \tag{4}$$

and embeds all 4D metrics in accordance with Campbell's theorem [5, 6]. The sign choice for the extra dimension is connected to the sign of $\Lambda$. This may be seen by considering the algebraically special 'pure' form of (4) or $C_5^*$ which has $g_{\alpha\beta} = g_{\alpha\beta}(x^\gamma$ only):

$$dS^2 = (\ell/L)ds^2 \pm d\ell^2 \tag{5.1}$$

$$ds^2 = g_{\alpha\beta}(x^\gamma)dx^\alpha dx^\beta \quad . \tag{5.2}$$

The Einstein field equations for this space reduce to the vacuum ones, which in terms of the 4D Ricci tensor are $R_{\alpha\beta} = \Lambda g_{\alpha\beta}$, with $\Lambda = \pm 3/L^2$ [2, 10]. Then in (5.1), the upper sign refers to $\Lambda < 0$ and the lower sign refers to $\Lambda > 0$. It is a corollary of Campbell's theorem that $C_5^*$ of (5) embeds all vacuum solutions of general relativity. And $C_5$ of (4) embeds all solutions with matter, the nature of the latter being specified by an induced or effective energy-momentum tensor whose form is now well known [2, 4]. In the present account, the focus is on dynamics, so (5) is relevant. The full equations of motion for (5) have been studied in detail; but it is not necessary for present purposes to repeat things, because geodesic paths are available directly from the metric (5) by taking $dS^2 = 0$ for the null-path. Indeed, it may be shown that the 5D null-path includes the 4D time-like path [4, 9]. This means that massive particles in 4D behave like photons in 5D.

The 4D paths $x^\gamma = x^\gamma(s)$ for (5) are in fact *identical* to the paths of massive particles in general relativity. This is because the canonical coordinates are specially chosen, so that the $(\ell/L)^2$ prefactor on the 4D part of the metric does not affect the motion as dictated by the 4D metric tensor $g_{\alpha\beta}$, which via $\partial g_{\alpha\beta}/\partial \ell = 0$ effectively decouples the 4D subspace from the 5D space. To better understand this 5D situation, two analogies are available from 4D Friedmann-Robertson-Walker cosmology. Firstly, the expanding Milne cosmology is isometric to Minkowski space, with a scale-factor proportional to the time $t$ which causes a factor $t^2$ to attach to the 3D subspace, in a fashion similar to how $\ell^2$ attaches to the 4D subspace in (5). Secondly, in all F.R.W. models, the spatial coordinates are commonly chosen to be comoving, so the time dependence is carried by the (square of the) scale-factor, in a fashion similar to how $\ell^2(s)$ carries the dependence on the proper time in (5). To state the situation in another way: the force (1) noted above, which exists in all 5D spaces that use the 4D proper time as parameter, is removed from the 4D subspace and confined to the 5D prefactor by choosing the coordinates in (5). This means that the 4D coordinate behaviour $x^\gamma(s)$ is the same as in general relativity, while the differences introduced by the extension of the geometry from 4D to 5D are con-fined to the behaviour $\ell(s)$ of the extra coordinate. Simplification of this kind only occurs in 5D mechanics for canonical coordinates.

Null-paths, with $dS^2 = 0$ but $ds^2 \neq 0$ in (5), result in two possible behaviours for the extra coordinate $x^4 = \ell$:

$$\ell = \ell_* e^{\pm s/L} \quad (\Lambda > 0) \tag{6.1}$$

$$\ell = \ell_* e^{\pm is/L} \quad (\Lambda < 0) \quad . \tag{6.2}$$

Here $\ell_*$ is a constant; and another such may be added to change the locus of the orbit, so the oscillation of (6.2) may be about $\ell = \ell_0$ rather than $\ell = 0$ if desired. Obviously, the monotonic motion (6.1) is the same in character as that noted previously for motion in $M_5$, and $\Lambda = 3/L^2$ here as before in (3). However, this identification for the 4D cosmological constant in terms of a 5D scale now has wider applicability, because the canonical metric (5) which leads to (6) applies to *any* embedded vacuum solution of general relativity.

Four-dimensional deSitter space is the simplest solution of general relativity to embed in $C_5^*$. Indeed, it can also be embedded in $M_5$ [7, 8]. In both contexts, it can be viewed as a pseudosphere of radius $L$. The 5D solution satisfies the Ricci-flat field equations $R_{AB} = 0$, which contain the 4D vacuum Einstein equations $R_{\alpha\beta} = \Lambda g_{\alpha\beta}$. With the inclusion of a mass $M$ at the centre of 3D space, the full 5D solution with a space-like extra coordinate is

$$dS^2 = \frac{\ell^2}{L^2}\left[\left(1-\frac{2M}{r}-\frac{r^2}{L^2}\right)dt^2 - \left(1-\frac{2M}{r}-\frac{r^2}{L^2}\right)^{-1}dr^2 - r^2(d\theta^2 + \sin^2\theta d\phi^2)\right] - d\ell^2 \quad . \tag{7}$$

In this, $\Lambda = 3/L^2$ as before. For $\Lambda > 0$, $\ell(s)$ is monotonic by (6.1). This is the case of most relevance, because modern observations indicate that the universe on the large scale is dominated by a fluid with the equation of state of the vacuum, which in the simplest interpretation is just the effect of the cosmological 'constant' with $\Lambda > 0$. However, for $\Lambda < 0$, $\ell(s)$ is oscillatory by (6.2), and this might be applied to the large vacuum fields of particle physics. Then, the wave can oscillate around $\ell = L$, so the fundamental mode has a wavelength equal to the circumference $2\pi L$. It is logical to identify this with the Compton wavelength $h/mc$ of the particle with mass $m$ associated with the wave. In terms of the local magnitude of the vacuum field as measured by $\Lambda$, this model leads to the relation $\Lambda = 3(mc/\hbar)^2$. While it may seem odd to derive results for both cosmological and particle physics in this way, the underlying reason is the same. Namely, the use of the *four*-dimensional proper time to describe a situation which is basically *five*-dimensional in nature.

As another application of this, brief mention should be made of Heisenberg's relation. This is commonly stated symbolically as $\Delta p \Delta x \geq \hbar$, and conventionally interpreted to mean that the motion and coordinates of a particle cannot both be measured to arbitrary accuracy. However, on dividing the noted relation by elements of the proper time, it involves in the appropriate limit the quantity $(dp/ds)(dx/ds)$. This describes a force parallel to the velocity, of the generic form (1) due to an extra dimension. It is therefore reasonable to ask if Heisenberg's relation is connected to the higher-dimensional force which remains after accounting for all other four-dimensional interactions. This question can be addressed by using either $M_5$ or $C_5$, since (2) and (6) give relationships between frames which both involve an exponential dependency on the 4D proper time $s$. Given this and preceding results, it is sufficient to note the main points of the argument. Thus the elements of displacement $dx^A$ and momentum change $dp_A$ may be combined to form the covariant quantity $dx^A dp_A$, and the result simplified by noting that the radius of the embedding deSitter pseudosphere is $L$ ($x^A x_A = L^2$). This on the basis of the previous model is related to the mass of the particle in the space by $L = \hbar/mc$. Then there comes $dx^A dp_A = (mc\,ds)^2/\hbar$. Assuming that the action is quantized, the smallest irremovable interaction in spacetime is thus characterized by $dx^\gamma dp_\gamma = \hbar$. (Along each axis of spacetime the answer is proportionately the same, because as shown in ref. 7 all points of deSitter space are equivalent.) In other words, Heisenberg's 4D relation has a form compatible with 5D dynamics, at least in a formal manner.

In the preceding, it was shown that when 4D spacetime is embedded in 5D, dynamical effects appear which resemble those conventionally attributed to the cosmological constant ($\Lambda$), Hubble's law (with $H$) and Heisenberg's relation (with $h$). The main reason for this is the use of 4D proper time ($ds$) in a space whose natural interval ($dS$) is different, causing an intrinsic mismatch of chronologies. This affects even flat 5D space ($M_5$), whose constant-velocity motion in 5D (2.1) becomes accelerated motion in 4D with a corresponding behaviour for the coordinates (2.2). In the static limit, the 5D and 4D coordinate times are given by (2.2) as $\tau = \tau_0 e^{\pm t/L}$ with suitably-defined constants. The acceleration involved is equivalent to the action of $\Lambda = 3/L^2$ by (3).

Generally, a force per unit mass or acceleration appears in spacetime if the 4D metric tensor depends on the extra coordinate $x^4 = \ell$ and there is relative motion between the 4D and 5D frames ($d\ell/ds \neq 0$). This force, given generically by (1), is parallel to the 4-velocity. However, a better way to embed 4D in 5D is to use not $M_5$ but canonical space $C_5$ as specified by (4). This metric is in general not 5D flat and not confined to describing vacuum. Many solutions of the 5D field equations with metric $C_5$ are known. When the 4D subspace does not depend on $x^4 = \ell$ except via the quadratic prefactor, the metric becomes the pure-canonical one $C_5^*$ of (5). This is important, because it is a theorem that $C_5^*$ can locally and smoothly embed any solution of the 4D Einstein equations with vacuum and a cosmological constant. Thus while 4D deSitter space can be embedded in $M_5$, 4D Schwarzschild space cannot and needs $C_5^*$ as in (7). It should be appreciated that the coordinates of $C_5$ are not the same as those of $M_5$ and in fact unique, being akin to the comoving ones of F.R.W. cosmologies. This is why it is impossible to test by purely dynamical means if the solar system is described by the 4D Schwarzschild solution or its 5D embedded form. The same comment applies to the global deSitter metric of inflationary cosmology. The 4D subspace of 5D $C_5^*$ space evolves in proportion to $\ell(s)$, whose behaviour is given by (6.1) for $\Lambda > 0$ and (6.2) for $\Lambda < 0$. The former choice is applicable to cosmology while the latter may be relevant to particle physics. It is because the essential physics is concentrated into the $\ell$-dependent prefactor in $C_5^*$ that an observer in the purely $x^\gamma$-dependent 4D subspace is unaware of the evolution in the 5D world. In particular, the force (1) due to the extra dimension goes unfelt in the embedded 4D spacetime. The wider implication of this is that if the universe is canonical 5D in nature, our physics may proceed in a 4D subspace, blithely unaware of the extra dimension.

3. Conclusion

The results in this note are really straightforward from the mathematical side; but from the physical side are significant because they concern the cosmological constant, Hubble's law and Heisenberg's relation. These things are all consistent with the hypothesis that 4D general relativity is embedded in a 5D space of canonical type.

By contrast, it is commonly assumed that if the world has more than 4 dimensions, then the embedding space is some version of Minkowski. A major motivation for this view is that any solution of Einstein's equations which is curved in 4 dimensions can be embedded in a manifold that is flat and has 10 or more dimensions. Hence supersymmetry and other approaches to unification. However, the present work and other recent developments [9, 11] show that 5D is completely adequate for mechanics, provided that the 4D subspace is given a quadratic dependency on the extra coordinate. It is plausible that the 'default' embedding manifold is not Minkowski space but rather canonical space.

Along similar lines, it is conceivable that the cosmological constant and related things are not merely *consistent* with canonical space, but have their *origin* there.

Acknowledgements

Comments came from members of the S.T.M. group, whose webpage is at http://astro.uwaterloo.ca/~wesson. Support was provided partly by N.S.E.R.C.

References

1. B. Mashhoon, P.S. Wesson, H. Liu, Gen. Rel. Grav. 30 (1998) 555.

2. P.S. Wesson, Space-Time-Matter, World Scientific, Singapore (2007), 197-210.

3. L. Randall, Science 296 (2002) 1422.

4. P.S. Wesson, Gen. Rel. Grav. 40 (2008) 1353.

5. S. Rippl., C. Romero, R. Tavakol, Class. Quant. Grav. 12 (1995) 2411.

6. S.S. Seahra, P.S. Wesson, Class. Quant. Grav. 20 (2003), 1321.

7. W. Rindler, Essential Relativity, Springer, New York, 2nd. ed. (1977) 185-188.

8. P.A.M. Dirac, Ann. Math. 36 (1935) 657.

9. P.S. Wesson, arXiv gr-qc/1011.0214 (2010).

10. H. Liu, B. Mashhoon, Ann. Phys. 4 (1995) 565. H. Liu, B. Mashhoon, Phys. Lett. A 272 (2000) 26.

11. M. Lachieze-Rey, Astron. Astrophys. 364 (2000) 894.